\documentclass[fleqn,10pt]{wlscirep}
\title{Lasing with cell-endogenous fluorophores: parameters and conditions}

\author[1,*]{Derrick Yong}
\author[1]{Ding Ding}
\affil[1]{Precision Measurements Group, Singapore Institute of Manufacturing Technology, 2 Fusionopolis Way, Innovis \#08-04, 138634, Singapore}

\affil[*]{derrick-yong@SIMTech.a-star.edu.sg}

\begin{abstract}
The notion of lasing with biologics has recently been realized and has since rapidly developed with the collective objective of creating lasers \textit{in vivo}. One limitation of achieving this goal is the requirement of exogenous laser dyes and fluorescent materials. To circumvent this, we investigate the use of cell-endogenous fluorophores - sources of cell autofluorescence - as laser gain material. In this work, we study the lasing potential and efficiency of flavins and reduced nicotinamide adenine dinucleotide (phosphate) (NAD(P)H) using a dye lasing model based on coupled rate equations. Analytical solutions for one- and two-photon pumped system were used in multi-parameter studies. We found that at physiological conditions, lasing can be supported by NAD(P)H with cavity quality factors of 10$^5$. With the further consideration of damage thresholds, we recommend the use of flavins as they entail lower threshold requirements. We then identify potential parameters for engineering to make the lasing of flavins feasible even at their low physiological intracellular concentrations. We also note the higher threshold requirements and lower efficiencies of two-photon pumping, but recognize its potential for realizing lasing \textit{in vivo}.
\end{abstract}
\begin{document}

\flushbottom
\maketitle

\thispagestyle{empty}

\section*{Introduction}
Biological lasers (bio-lasers) hold immense potential for applications within biological systems because they are themselves composed of biologics\cite{RN3283}. This concept of generating lasing within or by biologics would be able to circumvent the limited propagation of light in biological tissues as experienced by external laser sources. Since the first demonstration of a single-cell laser\cite{RN49} by Gather and Yun in 2011, biological lasers have developed rapidly in recent years. This included the demonstration of intracellular lasing using native cell organelles as microcavities\cite{RN51} for intracellular sensing as well as by internalizing microresonators\cite{RN51,RN53} for cell tagging and tracking. Aside from cells, bio-lasing has also been demonstrated with biomolecules (flavins\cite{RN52,RN3271}, green fluorescent protein\cite{RN50} and chloropyll\cite{RN46}) and human tissues (bone\cite{RN48} and blood\cite{RN47}). 

Albeit novel and remarkable demonstrations of biological lasers, the generation of lasing by biological cells and tissues still entail the use of externally introduced laser dyes or fluorescent material. Notably, there are biomolecules existing natively within cells that fluoresce. These cell-endogenous fluorophores are the source of autofluorescence, which is often regarded as noise in fluorescence microscopy. Notably, they are also the very machineries responsible for cell functions and metabolic activities\cite{RN1}. These biomolecules have therefore also been employed as endogenous biomarkers for applications like live cell characterization\cite{RN2,RN3} and cell sorting\cite{RN4,RN5}. Nevertheless, these are fluorescence emissions and are thus spectrally broad by nature. Such a property makes it difficult to discern between fluorophores with overlapping emission spectra. In contrast, lasing emissions are spectrally narrow and therefore would facilitate the differentation of emissions from several different fluorophores. In this work, we study the conditions and parameters for lasing two of the most abundant cell-endogenous fluorophores - flavins and reduced nicotinamide adenine dinucleotide (phosphate) (NAD(P)H). We do so by using and extending an established organic dye laser model\cite{RN12}. The framework of the model is based on coupled rate equations that describe the different energy states of our fluorophores. We analytically obtain the lasing thresholds and efficiencies for flavins and NAD(P)H, and identify parameters required for lasing under physiological conditions. We then make recommendations for possible approaches to lower threshold requirements so as to mitigate risks of inducing irreversible cell damage.

\section*{Methods} \label{methods}
\subsection*{Theoretical model}

\begin{figure}[ht]
\centering
\includegraphics[width=\linewidth*3/4]{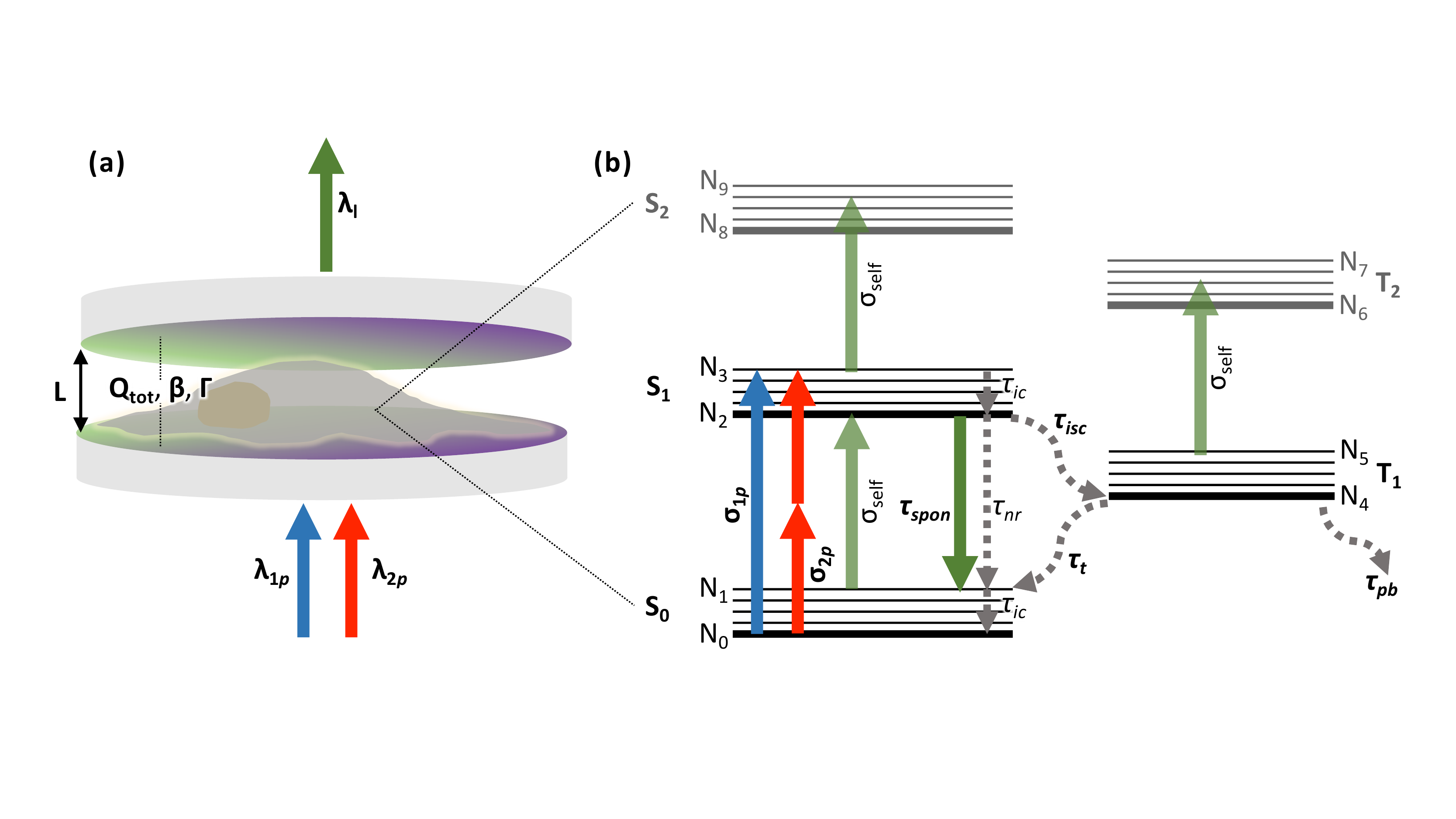}
\caption{Bio-laser system formed by an unlabelled cell. (a) Schematics of an adherent cell within an optical cavity. System is pumped either under a one- (in blue) or two-photon (in red) regime at the wavelengths $\lambda_{1p}$ and $\lambda_{2p}$ respectively. $\lambda_{l}$ is the lasing output (in green) wavelength. The cavity is defined by cavity length ($L$), total quality factor ($Q_{tot}$), spontaneous emission coupling factor ($\beta$) and confinement factor of lasing mode ($\Gamma$). (b) Energy level diagram for cell-endogenous fluorophores. N$_0$ to N$_9$ are the 10 energy levels of a fluorophore. The three lowest singlet states are marked as S$_0$, S$_1$ and S$_2$, while the two lowest triplet states are marked as T$_1$ and T$_2$. Absorption events are indicated by upward-pointing arrows and tagged with corresponding one-photon ($\sigma_{1p}$), two-photon ($\sigma_{2p}$) and self- ($\sigma_{self}$) absorption cross sections. Radiative and non-radiative relaxation events are denoted by downward-pointing solid and dotted arrows, correspondingly. Relaxation events are appended by their respective lifetimes for spontaneous emission ($\tau_{spon}$), non-radiative relaxation from S$_1$ to S$_0$ ($\tau_{nr}$), internal conversion ($\tau_{ic}$), intersystem crossing from S$_1$ to T$_1$ ($\tau_{isc}$), triplet relaxation from T$_1$ to S$_0$ ($\tau_{t}$) and photobleaching ($\tau_{pb}$).}
\label{fig:schematics}
\end{figure}

In this study, we investigated the parameters and conditions for lasing in biological cells without the introduction of exogenous laser dyes or fluorescent materials. We modelled the bio-laser construct as an adherent cell within an optical cavity, pumped by an external pulsed laser source as illustrated in Figure \ref{fig:schematics}a. The endogenous fluorophores, flavins and NAD(P)H, were analyzed for their potential as laser gain media under physiological conditions. It should be noted that flavins here refer to all three fluorescent derivatives found natively within cells, namely riboflavins (RF), flavin mononucleotides (FMN) and flavin adenine dinucleotide (FAD); while NAD(P)H collectively refers to the fluorescent reduced forms of nicotinamide adenine dinucleotide (NAD) and nicotinamide adenine dinucleotide phosphate (NADP). The parameters that define each endogenous fluorophore are listed in Table \ref{tab:systempara}. In the model, we assumed a homogenous distribution of either endogenous fluorophore within a 100$\mu$m$\times$100$\mu$m$\times$5$\mu$m volume. These dimensions correspond to the approximate dimensions of an adherent cell stretched over a 100$\mu$m$\times$100$\mu$m area with a height of 5$\mu$m -  equivalent to the cavity length.

The theoretical framework in this study is based on 11 coupled rate equations that describe 10 energy levels (N$_0$ to N$_9$) and an output as depicted by the energy level diagram in Figure \ref{fig:schematics}. These rate equations are based on an established organic dye laser model\cite{RN12}. To incorporate the two-photon pumping regime, the terms describing the rate of pumping were reformulated. In the one-photon regime it takes the form:
\begin{equation} \label{eq:rate_1p}
	rate_{1p} = \frac{I_{pump}A}{hf_{1p}}\times(1-10^{-\sigma_{1p}N_{den}L})
\end{equation}
Where $I_{pump}$ is the input pump intensity; $h$ is Planck's constant; $f_{1p}$ is the frequency of the input pump under one-photon pumping; $\sigma_{2p}$ is the one-photon absorption cross section; $N_{den}$ is the number density of the fluorophore, calculated from $N_AC/1000$ ($N_{A}$: Avogadro's constant; $C$: molar concentration); A is the area of pumping; and L is the thickness of the cavity. While in the two-photon regime, the rate of pumping takes the form:
\begin{align} \label{eq:rate_2p}
   rate_{2p} &= \frac{I^2_{pump}A}{2hf_{2p}}\times\frac{\sigma_{2p}N_{den}L}{1+I_{pump}\sigma_{2p}N_{den}L} \nonumber \\
   		  &\approx \frac{I^2_{pump}A}{2hf_{2p}}\times\sigma_{2p}N_{den}L
\end{align}
Where $f_{2p}$ is the frequency of the input pump under two-photon pumping and $\sigma_{2p}$ is its corresponding two-photon absorption cross section. Note that the approximation holds in this study across all considered fluorophore concentrations. Under the most ideal case of two-photon absorption by flavins at a concentration of 0.1M, deviation only occurs when pump intensities exceed 10$^{17}$Wcm$^{-2}$. All coupled rate equations were solved numerically in MATLAB. Parameters used in computations are listed in Table \ref{tab:systempara}.

\begin{table}[ht]
\centering
	\begin{tabular}{|l |c c|}
	\hline
	\textbf{Parameters}								& \textbf{Flavins}			& \textbf{NAD(P)H} \\
	\hline
	Pump area, $A$								& \multicolumn{2}{c|}{$1\times10^{-4}$cm$^{2}$} \\ 
	Cavity length, $L$								& \multicolumn{2}{c|}{5$\mu$m} \\ 
	Quality factor of cavity due to radiative loss, $Q_{rad}$	& \multicolumn{2}{c|}{$1\times10^5$} \\
	Spontaneous emission coupling factor, $\beta$			& \multicolumn{2}{c|}{$1\times10^{-4}$ $^($\cite{RN39}$^{,a)}$} \\
	Confinement factor of lasing mode, $\Gamma$			& \multicolumn{2}{c|}{0.2} \\
	
	One-photon pump wavelength, $\lambda_{1p}$			& \multicolumn{2}{c|}{390nm} \\
	Two-photon pump wavelength, $\lambda_{2p}$			& \multicolumn{2}{c|}{780nm} \\	
	Lasing output wavelength, $\lambda_{l}$				& 580nm 									& 500nm \\
	One-photon absorption cross section, 
	$\sigma_{1p}(\lambda_{1p})$						& $3.3\times10^{-17}$cm$^2$ $^($\cite{RN13}$^)$ 			& $2.0\times10^{-18}$cm$^2$ $^($\cite{RN14}$^)$ \\	
	Two-photon absorption cross section, 
	$\sigma_{2p}(\lambda_{2p})$						& $7.8\times10^{-33}$cm$^4$W$^{-1}$ $^($\cite{RN16}$^)$  	& $3.9\times10^{-35}$cm$^4$W$^{-1}$ $^($\cite{RN16}$^)$ \\
	Self-absorption cross section of output $S_0 \to S_1$, 
	$\sigma^{S_0S_1}_{self}$ 						& $10^{-20}$cm$^2$ 			 			& $10^{-21}$cm$^2$ \\		
	Self-absorption cross section of output $S_1 \to S_2$, 
	$\sigma^{S_1S_2}_{self}$ 						& $10^{-18}$cm$^2$ 			 			& $10^{-19}$cm$^2$ \\	
	Self-absorption cross section of output $T_1 \to T_2$, 
	$\sigma^{T_1T_2}_{self}$ 						& $10^{-18}$cm$^2$ 			 			& $10^{-19}$cm$^2$ \\	
												
	Fluorescence quantum yield, $\phi_{F}$				& 0.26 $^($\cite{RN18,RN19,RN20}$^)$ 			& 0.019 $^($\cite{RN21}$^)$  \\
	Spontaneous emission lifetime, $\tau_{spon}$			& 4.6ns $^($\cite{RN20,RN22,RN23}$^)$ 			& 0.4 $^($\cite{RN24,RN25,RN26,RN27}$^)$  \\
	Internal conversion lifetime, $\tau_{ic}$				& 1ps 									& 1ps  \\
	Intersystem crossing lifetime, $\tau_{isc}$				& 13.6ns $^($\cite{RN29}$^)$ 					& $\sim10^5$ $^($\cite{RN30}$^)$  \\
	Triplet decay lifetime, $\tau_{t}$						& 27$\mu$s $^($\cite{RN32}$^)$ 				& 2.7s $^($\cite{RN33}$^)$  \\

	Intracellular concentration, $C$						& $\sim10^{-6}$M	$^($\cite{RN45}$^{,b)}$		& $\sim10^{-5}$M $^($\cite{RN27}$^)$  \\	
	Critical transfer concentration, $C_0$				& $4.7\times10^{-2}$M $^($\cite{RN36}$^)$ 		& $3.5\times10^{-4}$M $^($\cite{RN37}$^)$ \\
	Dimerization constant, $K_D$						& 118M$^{-1}$ $^($\cite{RN38}$^)$ 				& NA $^($\cite{RN42,RN43}$^{,c)}$ \\
\hline
\end{tabular}
\caption{Parameters used in one- and two-photon pumped lasing models. $^{(a)}$Estimated based on the linewidth ratio between spontaneous emission and the lasing mode; $^{(b)}$Based on total flavin content per cell (i.e. combination of RF, FMN and FAD) and cell volume of ~$10^{-15}$m$^3$; $^{(c)}$No observations of NAD(P)H dimerization were reported, only electrochemically generated dimers of its non-fluorescent oxidized form (NAD(P)).}
\label{tab:systempara}
\end{table}

\subsection*{Lasing threshold}
Simplified analytical solutions to lasing thresholds were derived from the couple rate equations at steady state (i.e. $dN_i/dt$=0, where $N_{i}$ corresponds to the different energy levels). The total fluorophore population is assumed to be N$_{tot}$=N$_0$+N$_2$+N$_4$. This assumption is valid when: (i) pulsed excitation is considered where the rate of photobleaching ($1/\tau_{pb}$) is orders of magnitudes slower; (ii) pump absorption is negligible for S$_1$ to S$_2$ and T$_1$ to T$_2$ transitions; (iii) pump intensities are reasonably low, such that other levels are negligibly populated. The one-photon pumped lasing threshold\cite{RN12} is:
\begin{equation} \label{eq:Ithres_1p}
	I_{thres,1p} = \frac{hf_{1p}}{\sigma_{1p}N_{den}}\times\frac{{(\frac{1}{\phi_{F}\tau_{spon}}+\frac{1}{\tau_{isc}})/\tau^{S_0S_1}_{loss}}}{\frac{\beta\Gamma V}{\tau_{spon}}-(1+\frac{\tau_{t}}{\tau_{isc}})/\tau^{S_1S_2}_{loss}N_{den}}
\end{equation}
Where $\phi_{F}$ is the fluorescence quantum yield; $\tau_{spon}$, $\tau_{isc}$ and $\tau_{t}$ correspond to the lifetimes of spontaneous emission, intersystem crossing from S$_1$ to T$_1$ and triplet relaxation from T$_1$ to S$_0$ respectively; $\beta$ is the spontaneous emission coupling factor; $\Gamma$ is the confinement factor of the lasing mode; $V$ is the gain volume defined by $V=AL$; $\tau^{S_0S_1}_{loss}$ and $\tau^{S_1S_2}_{loss}$ are the combined losses from the passive cavity's photon decay lifetime ($tau_{loss}$) and self-absorption of the output from S$_0$ to S$_1$ and S$_1$ to S$_2$ (or T$_1$ to T$_2$) respectively. ($\tau^{S_0S_1}_{loss}=(1/\tau_{loss}+v_g\sigma^{S_0S_1}_{self}\Gamma N_{den})^{-1}$ and $\tau^{S_1S_2}_{loss}=(1/\tau_{loss}+v_g\sigma^{S_1S_2}_{self}\Gamma N_{den})^{-1}$, where $\tau_{loss}=Q_{tot}/2\pi f_l$; $f_l$ is the lasing frequency; $v_g$ is the group velocity of the lasing output; $\sigma_{self}$ is the self-absorption cross section). Similarly, we derived the lasing threshold under two-photon pumping as:
\begin{equation} \label{eq:Ithres_2p}
	I_{thres,2p} = \sqrt{\frac{2hf_{2p}}{\sigma_{2p}N_{den}}\times\frac{{(\frac{1}{\phi_{F}\tau_{spon}}+\frac{1}{\tau_{isc}})/\tau^{S_0S_1}_{loss}}}{\frac{\beta\Gamma V}{\tau_{spon}}-(1+\frac{\tau_{t}}{\tau_{isc}})/\tau^{S_1S_2}_{loss}N_{den}}}
\end{equation}
Effects of concentration quenching by dimerization\cite{RN4255,RN4254} on $\phi_F$ and $\tau_{spon}$ were also included in the lasing threshold analysis using the parameters acquired for the fluorophores' critical transfer concentration ($C_0$) and dimerization constant ($K_D$).  

\subsection*{Lasing efficiency}
Simplified analytical solutions to lasing efficiencies were likewise derived based on the same assumptions. The one-photon pumped lasing efficiency\cite{RN12} is:
\begin{equation} \label{eq:qlase_1p}
	q_{lase,1p} = \frac{f_l}{f_{1p}}\times\frac{(1-10^{-\sigma_{1p}N_{den}L})}{\tau_{cav}\Gamma}\times\frac{\frac{\beta\Gamma V}{\tau_{spon}}-(1+\frac{\tau_{t}}{\tau_{isc}})/\tau^{S_1S_2}_{loss}N_{den}}{\frac{\beta V}{\tau_{spon}\tau_{loss}}+v_g\sigma^{S_0S_1}_{self}[\frac{1}{\tau_{loss}}+(1+\frac{\tau_{t}}{\tau_{isc}})/\tau^{S_1S_2}_{loss}]}
\end{equation}
$q_{lase,1p}$ is essentially the gradient of the post-threshold linear slope for the input-output intensity plot (i.e. $q_{lase,1p}=dI_{out}/dI_{pump}$). Here, $\tau_{cav}$ is the photon decay lifetime due to radiative loss from the cavity. We also derive the two-photon pumped lasing efficiency as:
\begin{equation} \label{eq:qlase_2p}
	q_{lase,2p} = \frac{f_l}{2f_{2p}}\times\frac{\sigma_{2p}N_{den}L}{\tau_{cav}\Gamma}\times\frac{\frac{\beta\Gamma V}{\tau_{spon}}-(1+\frac{\tau_{t}}{\tau_{isc}})/\tau^{S_1S_2}_{loss}N_{den}}{\frac{\beta V}{\tau_{spon}\tau_{loss}}+v_g\sigma^{S_0S_1}_{self}[\frac{1}{\tau_{loss}}+(1+\frac{\tau_{t}}{\tau_{isc}})/\tau^{S_1S_2}_{loss}]}
\end{equation}
Where $q_{lase,2p}=dI_{out}/dI_{pump}^2$. Here, $q_{lase,2p}$ has been formulated as a dimensionless term like $q_{lase,1p}$.

\section*{Results}
\subsection*{Lasing thresholds}

\begin{figure}[ht]
\centering
\includegraphics[width=\linewidth*2/3]{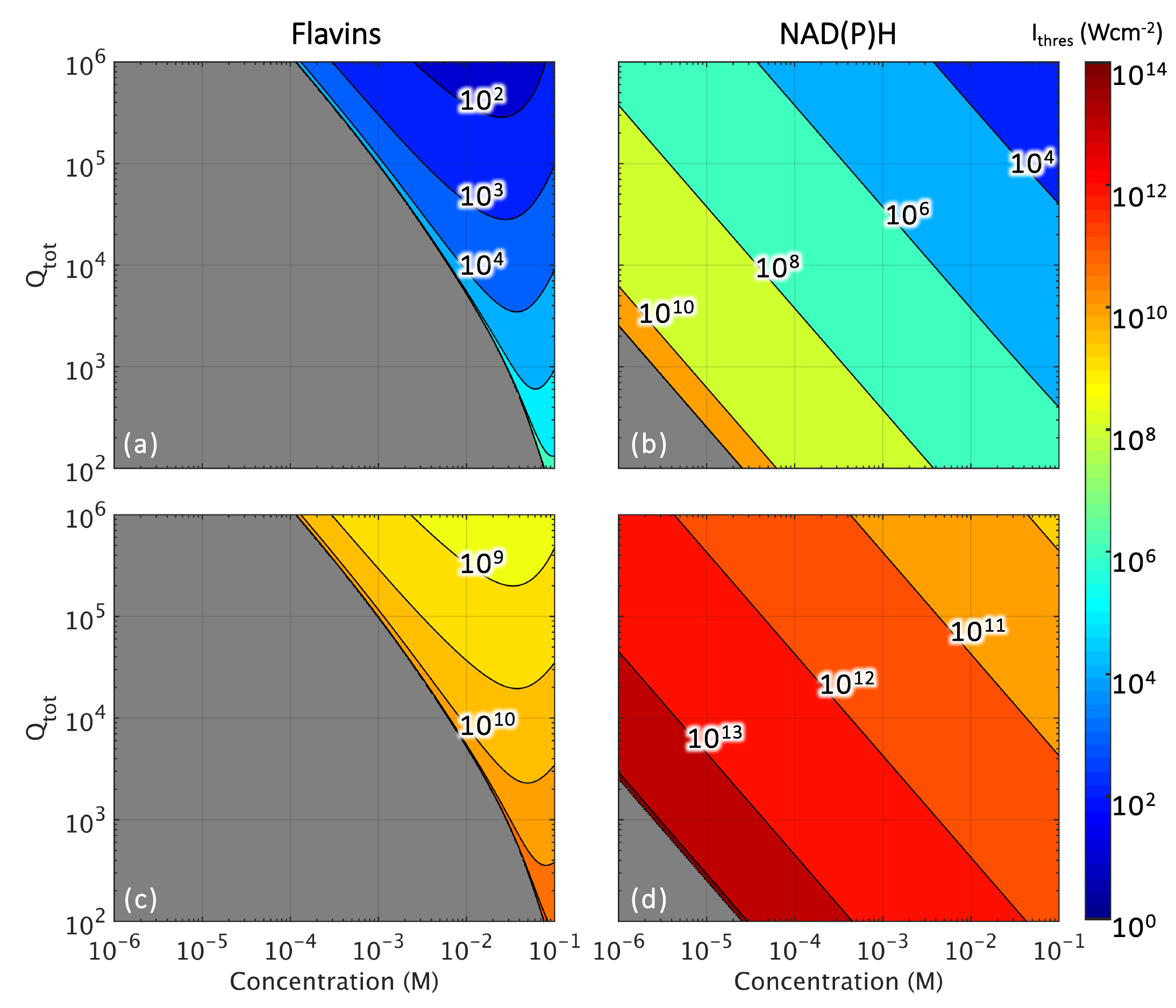}
\caption{Lasing threshold intensities (in Wcm$^{-2}$) as a function of total quality factor of the cavity ($Q_{tot}$) and fluorophore concentration ($C$) under (a,b) one- and (c,d) two-photon photon pumping for (a,c) flavins and (b,d) NAD(P)H. It should be noted that threshold intensity values have been plot based on their order of magnitude. Regions in grey indicate parameters that do not support lasing.}
\label{fig:Ithres_Qtot_conc_combi}
\end{figure}

In this section, we studied the order of pump intensities required for lasing by varying 3 parameters - total quality factor of cavity ($Q_{tot}$), concentration of fluorophores ($C$) and ratio of lifetimes for intersystem crossing ($\tau_{t}/\tau_{isc}$). Lasing thresholds were computed using Equations \ref{eq:rate_1p} and \ref{eq:rate_2p}. In Figure \ref{fig:Ithres_Qtot_conc_combi}, lasing thresholds are reported as a function of $Q_{tot}$ and $C$ and we note two key observations from the plots. First, NAD(P)H supports lasing over a wider range of concentrations as compared to flavins where lasing cannot be achieved below 10$^{-4}$M. This is true under both one- and two-photon pumping. Next, we note higher threshold intensities for NAD(P)H. Threshold intensities were computed to range from 10$^2$ to 10$^{11}$Wcm$^{-2}$ for NAD(P)H and 10$^1$ to 10$^6$Wcm$^{-2}$ for flavins under one-photon pumping. For two-photon pumping, threshold intensities ranged from 10$^9$ to 10$^{14}$Wcm$^{-2}$ for NAD(P)H and 10$^8$ to 10$^{11}$Wcm$^{-2}$ for flavins. 

\begin{figure}[ht]
\centering
\includegraphics[width=\linewidth*2/3]{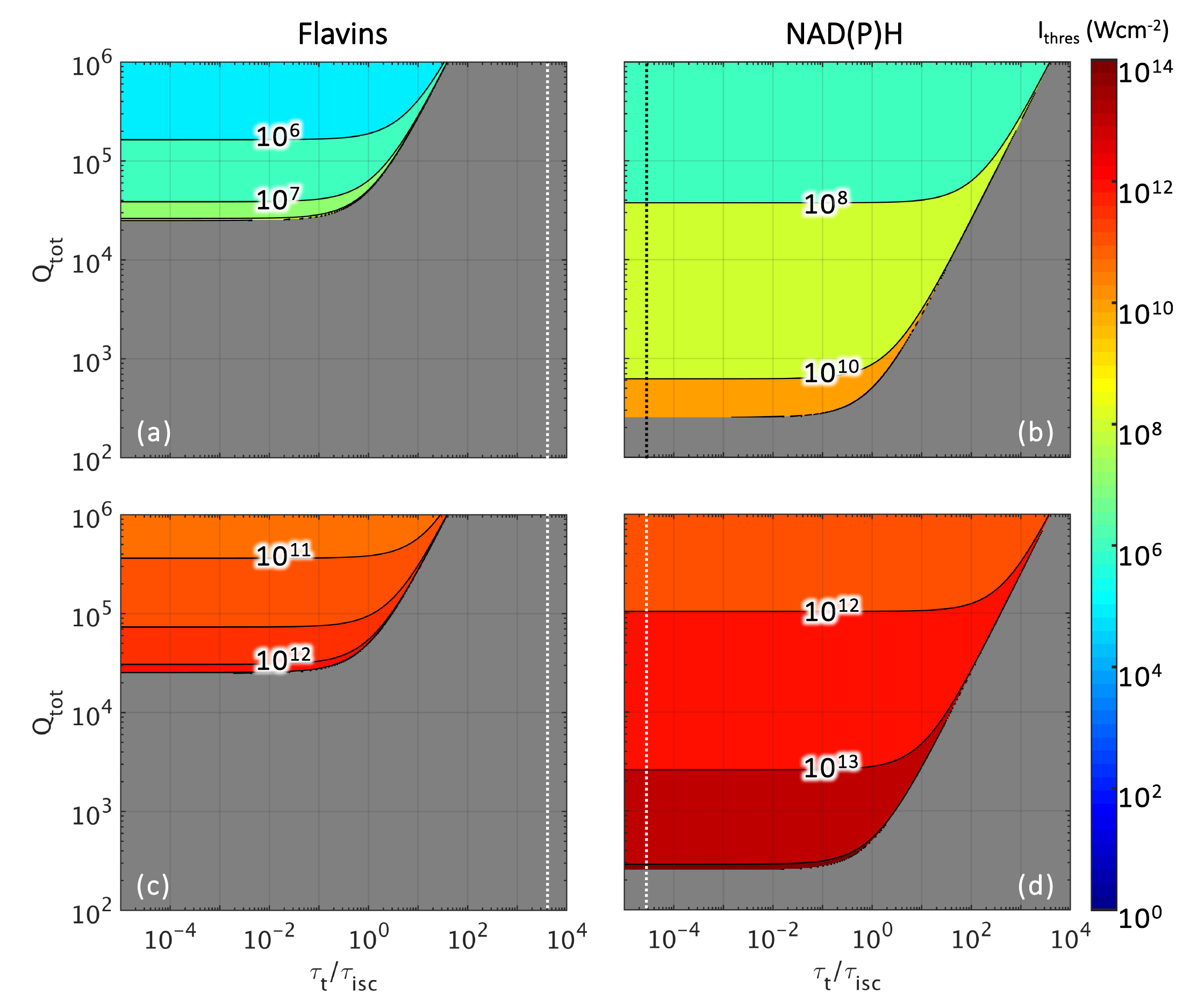}
\caption{Lasing threshold intensities (in Wcm$^{-2}$) as a function of total quality factor of the cavity ($Q_{tot}$) and ratio of lifetimes for intersystem crossing ($\tau_{t}/\tau_{isc}$) under (a,b) one- and (c,d) two-photon photon pumping for (a,c) 10$^{-6}$M flavins and (b,d) 10$^{-5}$M NAD(P)H. It should be noted that threshold intensity values have been plot based on their order of magnitude. Regions in grey indicate parameters that do not support lasing. Dotted lines denote $\tau_{t}/\tau_{isc}$ at physiological conditions ($\sim2\times10^{3}$ for flavins\cite{RN29,RN32} and $\sim3\times10^{-5}$ for NAD(P)H\cite{RN30,RN33}).}
\label{fig:Ithres_Qtot_tauttauisc_combi}
\end{figure}

In Figure \ref{fig:Ithres_Qtot_tauttauisc_combi}, we study the effects of varying $Q_{tot}$ and $\tau_{t}/\tau_{isc}$ on lasing thresholds. Threshold intensities were computed for intracellular concentrations of flavins\cite{RN45} and NAD(P)H\cite{RN27} at $\sim$10$^{-6}$M and $\sim$10$^{-5}$M respectively. Here, we note that lasing is supported in NAD(P)H at lower values of $Q_{tot}$. Under both pumping regimes, lasing can be achieved from $Q_{tot}$ of $2\times10^{2}$ for NAD(P)H and $2\times10^{4}$ for flavins. In addition, we highlight that at physiological conditions of  $\tau_{t}/\tau_{isc}$, lasing is not supported by flavins. This is indicated by the white-dotted line residing within the non-lasing region of the Figure \ref{fig:Ithres_Qtot_tauttauisc_combi}a and c.

\subsection*{Lasing efficiencies}

\begin{table}[ht]
\centering
	\begin{tabular}{|l |c c c|}
	\hline
					& \multicolumn{3}{c|}{$Q_{tot}$} \\	
		 			& $1\times10^{5}$					& $1\times10^{4}$					& $1\times10^{3}$ \\
	\hline
	$q_{lase,1p}$ 		& $1.05\times10^{-5}$				& $1.03\times10^{-6}$				& $7.88\times10^{-8}$ \\
	$q_{lase,2p}$ 		& $9.20\times10^{-23}$				& $8.97\times10^{-24}$				& $6.91\times10^{-25}$ \\	
	\hline
\end{tabular}
\caption{Lasing efficiencies for $1\times10^{-5}$M NAD(P)H under one- and two-photon pumping in systems with different total cavity quality factor ($Q_{tot}$).}
\label{tab:qlase}
\end{table}

\begin{figure}[ht]
\centering
\includegraphics[width=\linewidth/2]{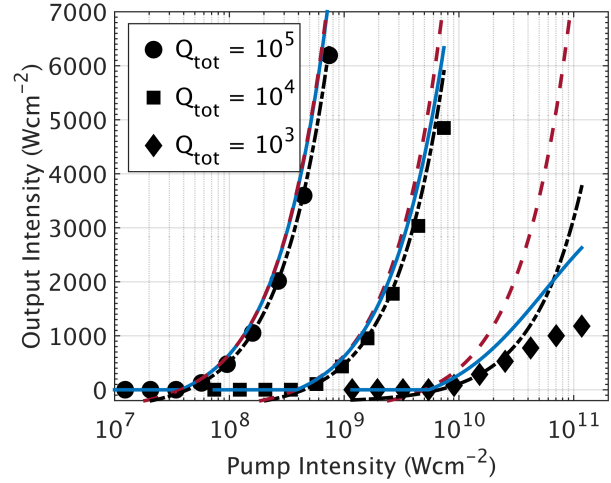}
\caption{Output-input intensity plots for one-photon pumped $1\times10^{-5}$M NADH in systems with different total cavity quality factor ($Q_{tot}$). Numerically computed data are represented by solid black circles, squares and diamonds for $Q_{tot}$ of 10$^5$,10$^4$ and 10$^3$ respectively. Black dash-dotted lines are linear fits of the first two numerical data points post-lasing threshold. Red dashed lines are plots obtained from the simplified analytical solutions to lasing threshold ($I_{thres,1p}$) and efficiency ($q_{lase,1p}$). Blue lines are semi-simplified analytical solutions to the coupled rate equations.}
\label{fig:qlase_1p}
\end{figure}
\begin{table}[ht]
\centering
	\begin{tabular}{|l |c c c|}
	\hline
												& \multicolumn{3}{c|}{$Q_{tot}$} \\	
		 										& $1\times10^{5}$		& $1\times10^{4}$		& $1\times10^{3}$ \\
	\hline
	$q_{lase,1p}$ from simplified analytical solution			& 15.7$\%$			& 21.4$\%$			& 132$\%$ \\
	Linear fit of semi-simplified analytical solution			& 15.4$\%$			& 18.2$\%$			& 58.5$\%$ \\
	\hline
\end{tabular}
\caption{Percentage deviation of analytically-obtained lasing efficiencies from numerical data for one-photon pumped $1\times10^{-5}$M NAD(P)H in systems with different total cavity quality factor ($Q_{tot}$).}
\label{tab:numanaldata_qlase1p}
\end{table}

We further analyzed the efficiency of lasing for 10$^{-5}$M NAD(P)H for different $Q_{tot}$. Lasing efficiency corresponds to the amount of output lasing intensity generated per unit of input pump intensity (or square of input pump intensity for the two-photon regime) . Efficiencies were computed using Equations \ref{eq:qlase_1p} and \ref{eq:qlase_2p} and are reported in Table \ref{tab:qlase}. For both regimes of pumping, efficiencies are noted to decrease by approximately an order per order of decrease in $Q_{tot}$. We further observe that efficiencies under one-photon pumping are 18 to 19 orders higher than their two-photon pumped counterparts.

Next, we examined the deviation of simplified analytical solution from numerically computed results. Initial computations revealed significant deviations from the numerical solution especially at low $Q_{tot}$. At high pump intensities, assumption (iii) that was stated in the derivation of lasing threshold solutions becomes invalid. 
A semi-simplified analytical solution was thus derived by considering N$_{tot}$=N$_0$+N$_2$+N$_3$+N$_4$. This solution under one-photon pumping was derived as follows. First we obtained expressions to eliminate terms defining the populations at different energy levels:
\begin{equation} \label{eq:F30}
	F_{30} =  \frac{N_3}{N_0} = \frac{\frac{I_{pump}A}{hf_{1p}N_{tot}}(1-10^{-\sigma_{1p}N_{den}L})}{\frac{1}{\tau_{ic}}+\frac{I_{pump}A}{hf_{1p}N_{tot}}(1-10^{-\sigma_{1p}N_{den}L})}
\end{equation}
\begin{equation} \label{eq:F02}
	F_{02} =  \frac{N_0}{N_2} = \frac{\frac{I_{out}A}{hf_l}(v_g\sigma^{S_0S_1}_{self}\Gamma +\frac{\beta\Gamma}{\tau_{spon}})+(\frac{1}{\phi_{F}\tau_{spon}}+\frac{1}{\tau_{isc}})}{\frac{I_{pump}A}{hf_{1p}N_{tot}}(1-10^{-\sigma_{1p}N_{den}L})(1-F_{30})+\frac{I_{out}A}{hf_lV}(v_g\sigma^{S_0S_1}_{self}\Gamma)}
\end{equation}
The above expressions were then incorporated into the following equation:
\begin{equation} \label{eq:semianalsoln}
	\frac{I_{pump}A}{hf_{1p}}\times(1-10^{-\sigma_{1p}N_{den}L})\times\frac{F_{02}(1-F_{30})}{F_{02}(1+F_{30})+(1+\frac{\tau_{t}}{\tau_{isc}})}  = \frac{N_{tot}(\frac{1}{\phi_{F}\tau_{spon}}+\frac{1}{\tau_{isc}})}{F_{02}(1+F_{30})+(1+\frac{\tau_{t}}{\tau_{isc}})}+\frac{I_{out}A}{hf_l\tau_{loss}}
\end{equation}
A semi-simplified analytical solution for I$_{out}$ was then solved from the above using MATLAB. The same was done for the two-photon regime, by replacing the rate of pumping terms from what is defined in Equation \ref{eq:Ithres_1p} to that in Equation \ref{eq:Ithres_2p}. 

In Figure \ref{fig:qlase_1p}, we see the differences between the simplified (red dashed lines) and semi-simplified analytical solutions (blue solid lines) and the numerical data (black-coloured data points). This difference is observed to increase with decreasing $Q_{tot}$. This difference is further quantified by taking the difference between analytical and numerical solutions as a percentage of the corresponding numerical solutions. We report these calculated percentage deviations for lasing efficiency in Table \ref{tab:numanaldata_qlase1p}. To obtain lasing efficiencies from the numerical solution, linear fits were made with the first two data points post-lasing threshold. The corresponding range of data points for the semi-simplified analytical solution were then used in obtaining their respective lasing efficiencies.

\section*{Discussion} 
\subsection*{Lasing at physiological conditions}
At physiological conditions, where intracellular concentrations of flavins and NAD(P)H are at $\sim$10$^{-6}$M and $\sim$10$^{-5}$M respectively, we note from Figure \ref{fig:Ithres_Qtot_conc_combi} that it is not possible to lase flavins. It is however possible to lase NAD(P)H with high pump intensities. According to the ICNIRP guidelines\cite{RN4092}, the damage threshold for tissues (skin) exposed to visible lasers of nanosecond pulses is 20mJcm$^{-2}$ - for a 1ns pulse, this is equivalent to 20MWcm$^{-2}$. From Figure \ref{fig:Ithres_Qtot_conc_combi}b, we identify a required total cavity quality factor of $Q_{tot}>$10$^5$ to lase NAD(P)H without inducing damage. Such quality factors are attainable by whispering gallery mode (WGM) microresonators\cite{RN3973} and even Fabry-P\'{e}rot microcavities that enable three-dimensional confinement\cite{PhysRevB.87.161116}. However, considering that a WGM microresonator requires the fluorophores to reside at its periphery to effectively generate lasing, it becomes a challenge for microresonators in a cell to be able to tap on all of the intracellular fluorophores. Fabry-P\'{e}rot cavities, on the contrary, offer a more efficient means of utilizing all of the cell's contents since the fluorophore need only be sandwiched between a pair of mirrors. On the other hand, when pumped at NIR wavelengths, the ICNIRP guidelines do not state any thresholds for sub-nanosecond pulses for irradiation of skin tissue, instead it only suggests a conservative damage threshold of $\sim$20GWcm$^{-2}$ for NIR femtosecond pulses. The guidelines do however list a damage threshold of 1MWcm$^{-2}$ for exposure of the eye to 100fs NIR pulses. A study\cite{RN4475} on cells in culture irradiated by multiple (76MHz over 0.25s) 130fs NIR (810nm) pulses at a beam diameter of $\sim$100$\mu$m reports a damage threshold of 1.9kJcm$^{-2}$. By taking the very conservative assumption that damage is induced upon irradiation by the very first pulse (at 0.1mJcm$^{-2}$), we estimate a threshold intensity of $\sim$1GWcm$^{-2}$. From Figure \ref{fig:Ithres_Qtot_conc_combi}d, we see that lasing NAD(P)H is possible at physiological conditions under two-photon pumping by NIR lasers, but the reported damage thresholds imply the cell's inevitable demise. It is hence more feasible to lase NAD(P)H in cells under the one-photon regime than by its two-photon counterpart. 

It is interesting to note that although NAD(P)H have both absorption cross sections and quantum yields at an order of magnitude smaller than flavins, they are able to support a much larger range of conditions for lasing. This is counterintuitive considering that flavins are able to absorb and emit more efficiently than NAD(P)H. From the energy level diagram in Figure \ref{fig:schematics}b, we understand this to be due to intersystem crossing from the singlet (S$_1$) to triplet (T$_1$) state. In flavins, the intersystem crossing lifetime\cite{RN29} ($\tau_{isc}$) is of the same nanosecond timescale as its spontaneous emission lifetime\cite{RN20,RN22,RN23} ($\tau_{spon}$). This implies competition between transitions from the upper laser level (N$_2$) to either the lower laser level (N$_1$) or the lower triplet level (N$_4$). Furthermore, the decay rate of flavins from its triplet to singlet state is three orders slower than its rate of triplet formation, resulting in the ``trapping'' of excited fluorophores in a energy state not usable in lasing transitions. In comparison, NAD(P)H has virtually negligible triplet transitions\cite{RN30}. From Figure \ref{fig:Ithres_Qtot_tauttauisc_combi}a and c, we observe again that at physiological conditions (denoted by vertical dotted lines) lasing is not supported by flavins. From the plots, we also see that reducing $\tau_{t}/\tau_{isc}$ to a value of $\sim$10 would render flavins capable of lasing at its physiological intracellular concentration. This can be achieved by reducing $\tau_{t}$ and increasing $\tau_{isc}$, which translates to faster decays from the triplet state and reduced intersystem crossings respectively. The use of iodide as quenchers has been reported to reduce the triplet population and increase fluorescence in flavins when used in small quantities\cite{VANDENBERG20012135}. Although iodide is known to increase the rate of intersystem crossing (results in higher triplet populations), the authors have attributed their findings to the concomitant increase in triplet decay rates, which undergo a relatively higher increment than that of intersystem crossing. Alternatively, binding of flavins with light-oxygen-voltage-sensing (LOV) proteins has also been demonstrated to reduce the rate of intersystem crossing\cite{RN29}, again resulting in a lower triplet population. Considering the several-order lower lasing threshold intensities required by flavins, we recommend that an approach involving the engineering of the above-mentioned lifetimes be adopted. When $\tau_{t}/\tau_{isc}<$1, lasing is achievable in flavins at physiological intracellular concentrations with just a $Q_{tot}$ of $\sim$4$\times$10$^4$ and by pump intensities under the damage threshold. Such a $Q_{tot}$ is still attainable under the most ideal conditions of a Fabry-P\'{e}rot cavity comprising a mirror pair. 

\subsection*{One- vs. two-photon pumping}
When we compare one- and two-photon pumping, we note two keys points for discussion. First, threshold intensity requirements are many orders of magnitude higher for the two-photon regime. This is consequent of the nature of multi-photon processes, where two or more photons have to arrive simultaneously for energy transitions to take place. Such a requirement translates into a large number of photons arriving over a short period of time in a small area, ergo high intensities (in Wcm$^{-2}$). That being said, it should be highlighted that such a property is in fact an advantage of multi-photon processes, particularly in the biosciences\cite{RN10}. The high intensity requirements enable spatial specificity and transparency of materials at low intensities. Although not considered in this study, this property could enable the lasing of specific cells buried within tissues provided that cavity requirements can be met intracellularly. Furthermore, the typical use of NIR wavelengths in such process also enables deeper penetration \textit{in vivo}\cite{RN11}, indicating the possibility of \textit{in vivo} laser generation. Similarly, as would be expected, lasing efficiencies are many orders of magnitude lower considering the high intensities already required to first excite the fluorophores. Secondly, we note the parameters that do not support lasing are dependent only on the fluorophore and not the regime of pumping. This is seen in the identical grey regions shaded in Figures \ref{fig:Ithres_Qtot_conc_combi} and \ref{fig:Ithres_Qtot_tauttauisc_combi}. From Equations \ref{eq:Ithres_1p} and \ref{eq:Ithres_2p}, lasing is not supported when the denominator of the last collection of terms becomes negative. This occurs when $\beta\Gamma V/\tau_{spon} < (1+\tau_{t}/\tau_{isc})/\tau^{S_1S_2}_{loss}N_{den}$ (refer to Methods and Table \ref{tab:systempara} for details) - i.e. when the maximal available gain is smaller than the intrinsic losses of the system. This is thus independent on the pump but still dependent on cavity properties, namely the spontaneous emission coupling factor ($\beta$) and lasing mode confinement ($\Gamma$). However, should the spatial specificity of multi-photon processes be considered, the pump area effectively changes and consequently the available fluorophores within said area. It is thus particularly important to note the beam diameter when employing the two-photon regime described in this study. 

\subsection*{Saturation of higher excited state (N$_3$)}
At high pump intensities we observe a deviation the simplified analytical solution from its numerical counterpart. We attribute this to the breakdown of the assumption that only N$_0$, N$_2$ and N$_4$ are populated. When pumped at high intensities, the rate of internal conversion from N$_3$ to N$_2$ ($1/\tau_{ic}=1\times10^{12}s^{-1}$) becomes insufficient to significantly depopulate N$_3$. This results in saturation of the fluorophore population in this higher excited state, which is different from the typical saturation we observe in two-level laser systems. For 10$^{-5}$M NAD(P)H, the rate of excitation from N$_0$ to N$_3$ matches the rate of depopulation from N$_3$ to N$_2$ when pump intensities exceed 10$^4$Wcm$^{-2}$ and 10$^{10}$Wcm$^{-2}$ under one- and two-photon pumping, correspondingly. From threshold intensities in Figure \ref{fig:Ithres_Qtot_conc_combi}, we note that this applies to all ranges of parameters studied. Hence, non-negligible populations of fluorophores exist in N$_3$. A semi-simplified analytical solution was hence derived by considering N$_{tot}$=N$_0$+N$_2$+N$_3$+N$_4$. We compared the simplified analytical, semi-simplified analytical and numerical solutions in Figure \ref{fig:qlase_1p} and further quantified deviations of the analytical solutions from the numerical solution in Table \ref{tab:numanaldata_qlase1p}. It should be noted that the x-axis of the input-output plot is in the log scale, which accounts for the non-linear curves. We first observe that all lasing thresholds were well within the same order of magnitude, implying no need to derive more complex solutions for threshold computations. Lasing efficiencies, on the other hand, are observed to differ significantly in Figure \ref{fig:qlase_1p}, especially for a low $Q_{tot}$ of 10$^3$. We also see how the semi-simplified analytical solution better matches the numerical solution, with its relative amount of deviation improving for lower values of $Q_{tot}$. In that regard, we recommend the use of the semi-simplified analytical solution when studying poorly performing cavities so as to not get an over estimate of lasing efficiency. 

\section*{Conclusion} \label{conclusion}
In summary, we have theoretically studied the feasibility of lasing with cell-endogenous fluorophores and identified parameters that allow for lasing in cells at physiological conditions. We found that lasing is supported by NAD(P)H but not flavins at physiological intracellular concentrations. This could be achieved using cavities with $Q_{tot}>$10$^5$ under the one-photon pumping regime. From further analysis of the fluorophores's transitions, we identified the intersystem crossing in flavins to be a key reason for its inability to support lasing. Recommendations on methods of engineering $\tau_{t}$ and $\tau_{isc}$ were made so as to allow lasing of flavins to be achievable at physiological intracellular concentrations. In short, lasing flavins would be preferred over NAD(P)H due to its lower threshold requirements, which implies lower risks of inducing damage to cells. We also highlight the benefits of lasing under the two-photon regime, which we recognize to hold more potential for \textit{in vivo} applications than the one-photon regime. In conclusion, we summarize from this theoretical study that lasing unlabelled cells is possible, and can be further developed with novel methods of (i) integrating high quality factor optical cavities with cells; and (ii) minimizing intersystem crossings in fluorophores intracellularly.


\section*{Acknowledgements}

This work was supported by the Agency for Science Technology and Research (A$^*$STAR), Singapore.

\section*{Author contributions statement}

D.Y. conceived the idea, D.Y. formulated the algorithms and conducted the computations, D.Y. and D.D. studied and refined the model, D.Y. and D.D. analysed and articulated the results. All authors reviewed the manuscript. 

\section*{Additional information}
\textbf{Competing financial interests} The authors declare no competing financial interests. 



\end{document}